\documentclass[%
 reprint,
 amsmath,amssymb,
 aps,
]{revtex4-2}

\usepackage{graphicx}
\usepackage{dcolumn}
\usepackage{bm}

\usepackage{titlesec}

\titleformat{\section}{\centering\normalsize\bfseries}{\thesection}{1em}{}

\titleformat{\subsection}{\centering\normalsize\bfseries}{\thesubsection}{1em}{}

\titleformat{\subsubsection}{\centering\normalsize\bfseries}{\thesubsubsection}{1em}{}

\renewcommand\thesection{\Roman{section}.}
\renewcommand\thesubsection{\Alph{subsection}.}
\renewcommand\thesubsubsection{\thesubsection\arabic{subsubsection}.}

\usepackage{indentfirst}
\usepackage{ragged2e,etoolbox}

\usepackage[caption=false]{subfig}   
\usepackage{caption}                 

\DeclareCaptionJustification{just}{\justifying}
\usepackage{amsmath}
\usepackage[utf8]{inputenc}
\usepackage[utf8]{inputenc}
\usepackage{booktabs}
\usepackage[colorlinks=true, linkcolor=blue, citecolor=blue, urlcolor=blue]{hyperref}
\usepackage{overpic}
\usepackage[utf8]{inputenc}
\usepackage{textgreek}
\usepackage{tabularx}
\usepackage{verbatim}
\usepackage{soul}

\makeatletter
\renewcommand\thesection{\Roman{section}} 
\makeatother

\begin{document}

\preprint{APS/123-QED}

\title{Exploring the role of higher $\omega$ meson states in the $e^+ e^-\rightarrow b_1(1235) \pi$ process}

\author{Zhao-Yang Wu$^{1}$}
\author{Zi-Yue Bai$^{2,3,4}$}\email{baiziyue@lzu.edu.cn}
\author{Li-Ming Wang$^{1}$\footnote{Corresponding author}}\email{lmwang@ysu.edu.cn}
\affiliation{ $^1$Key Laboratory for Microstructural Material Physics of Hebei Province, School of Science, Yanshan University, Qinhuangdao 066004, China\\
$^2$School of Physical Science and Technology, Lanzhou University, Lanzhou 730000, China\\
$^3$5Lanzhou Center for Theoretical Physics, Key Laboratory of Theoretical Physics of Gansu Province,
 Key Laboratory of Quantum Theory and Applications of MoE, Gansu Provincial Research Center
 for Basic Disciplines of Quantum Physics, Lanzhou University, Lanzhou 730000, China\\
$4$Research Center for Hadron and CSR Physics, Lanzhou University and Institute of Modern Physics of CAS, Lanzhou 730000,China
}

\date{\today}

\begin{abstract}
The properties of light vector mesons near 2.2 GeV remain poorly understood, impeding progress in mapping the higher-lying hadronic spectrum. Utilizing the newly released BESIII data on the Born cross sections for the process of $e^+ e^- \rightarrow b_1(1235) \pi$, we conduct a combined analysis incorporating theoretical predictions for the mass spectrum and decay properties of $\omega$-meson family. Our fit demonstrates that the enhancement structure near 2.2 GeV originates not from a single resonance, but from the significant interference between the $\omega(4S)$ and $\omega(3D)$ states, which have comparable contributions. This interpretation resolves the apparent discrepancy in the resonance parameters and yields values consistent with theoretical expectations.  Our work provides a key interpretation of  the vector enhancement structure and establishes a vital framework for identifying higher radial and orbital excitations in the $\omega$ meson family, thereby advancing the mapping of the light-hadron spectrum.

\end{abstract}

\maketitle


\section{\label{sec:level1}INTRODUCTION}
\vspace{2ex}

Hadron spectroscopy provides a crucial window into the non-perturbative regime of Quantum Chromodynamics (QCD). Significant progress in this field, particularly in mapping the light-flavor meson spectrum, has been driven by advances in experimental capabilities. Recent observations of numerous new hadronic states in $e^+e^-$ annihilation data at center-of-mass energies around $\sqrt{s}\sim$ 2 GeV \cite{BaBar:2019kds,BESIII:2020xmw,BESIII:2021bjn,BESIII:2020kpr,BESIII:2021uni,BESIII:2021aet,BESIII:2022yzp} have further intensified research interest in the higher-lying states of this spectrum.

A key recent development is the measurement by the BESIII Collaboration of the Born cross section fo $e^+ e^- \rightarrow b_1(1235) \pi$ in the energy range from 2.0 to 3.08 GeV \cite{BESIII:2022yzp}. Their analysis revealed a distinct resonance with a mass of  $M = 2200 \pm11\pm17$ MeV and a width of  $\Gamma=74\pm20\pm24$ MeV,  exhibiting a combined statistical significance of $7.9\sigma$. This observation provides a new and precise experimental anchor point for theoretical investigations.

\begin{figure}[hbp]
  \centering
  \includegraphics[width=\columnwidth]{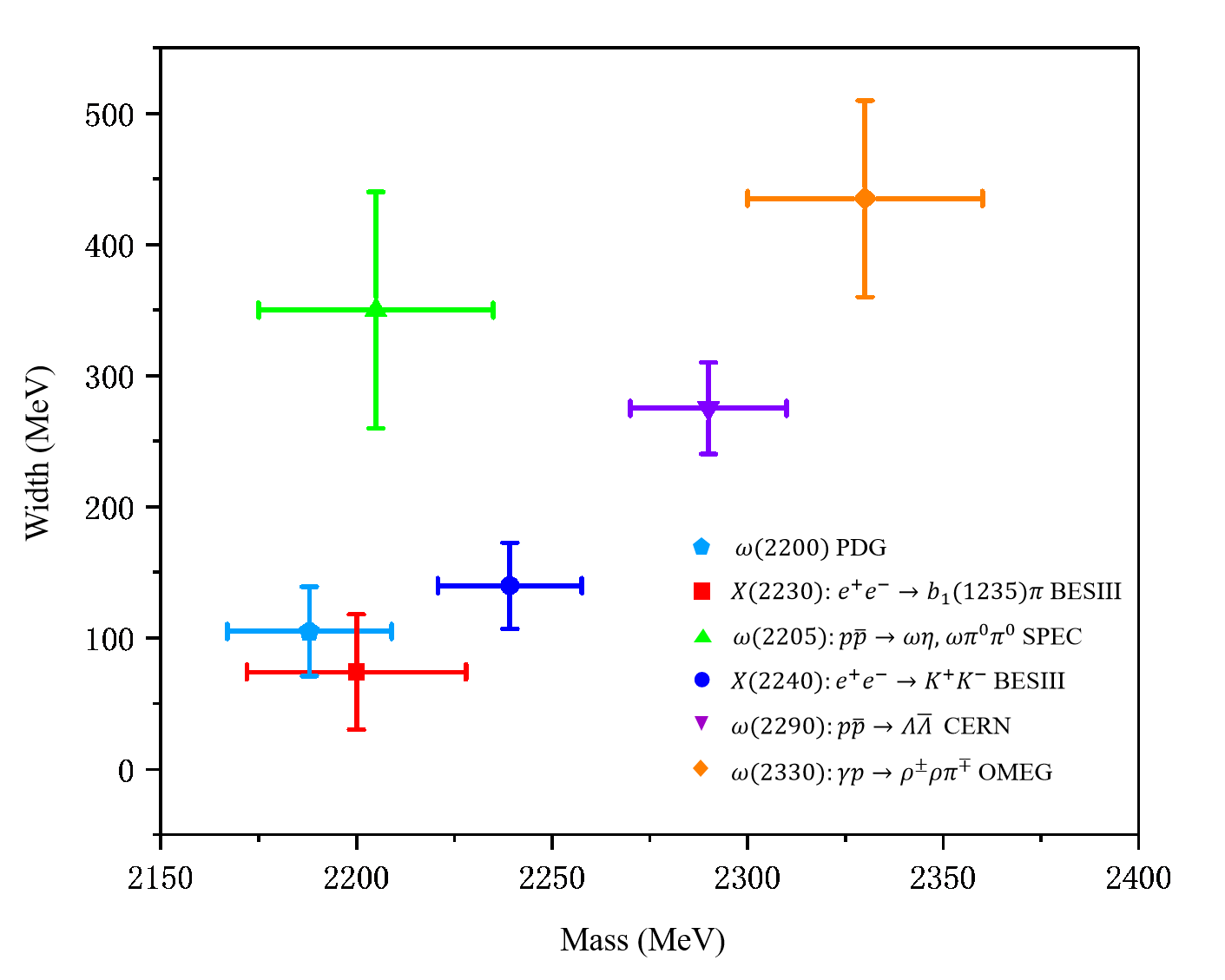} 
  \caption{A comparison of resonance parameters of these reported $\omega$ states with masses around 2.2 GeV ~\cite{Anisovich:2002xoo,Bugg:2004rj,hepdata.33892,BESIII:2018ldc}}
  \label{fig:masses around 2.2}
\end{figure}

The interpretation of such data relies on a clear understanding of the established meson spectrum. 
The established $\omega$ meson spectrum below 2 GeV includes the $\omega(782)$ as the $S$-wave ground state, $\omega(1420)$ and $\omega(1960)$ as candidate $2S$ and $3S$ radial excitations, and $\omega(1650)$ as the $D$-wave ground state \cite{ParticleDataGroup:2024cfk,Pang:2019ovr,Barnes:1996ff,Ebert:2005ha,Ebert:2009ub,Wang:2012wa}. However, the situation above 2 GeV becomes considerably more complex in the 2.0-2.3 GeV region, where several $\omega-$like states have been reported, leading to ongoing debates about their internal structure and assignments. For instance, the Lanzhou group~\cite{Pang:2019ovr} proposed identifying the $\omega(2290)$ and $\omega(2330)$ as the $\omega(4S)$ state and $\omega(2205)$ as the $\omega(3D)$ state. Despite these proposals, persistent discrepancies between predicted and measured properties, as well as the uncertain status of other candidates (as seen in Fig.~\ref{fig:masses around 2.2}), highlight the incomplete and ambiguous nature of the $\omega$ spectrum above 2 GeV. This underscores the necessity for more precise data to clarify the properties and verify the existence of these states.

The recent high-precision data from BESIII on $e^+ e^- \rightarrow b_1(1235) \pi$ \cite{BESIII:2022yzp} thus offer a critical opportunity to test specific theoretical assignments, particularly for the $\omega(4S)$ and $\omega(3D)$ states. In this work, we perform a comprehensive analysis of this process by incorporating theoretical input from $\omega$ meson spectroscopy. Our results demonstrate that the observed enhancement near 2.2 GeV can be explained by the significant contributions of both the $\omega(4S)$ and $\omega(3D)$states, which are found to have comparable influence in this specific channel.

The paper is structured as follows. Following this Introduction, Sec.~\ref{sec:level2} details the theoretical analysis of the $\omega$ meson family around 2 GeV, including the mass spectrum and decay behaviors. Sec.~\ref{sec:level3} presents the framework for calculating the cross section of $e^+ e^- \rightarrow b_1(1235) \pi$. In Sec.~\ref{sec:level4}, we perform a fit to the BESIII data to extract the contributions of the $\omega(4S)$ and $\omega(3D)$ resonances. Finally, we summarize our findings and discuss their implications in Sec.~\ref{sec:level5}.

\setlength{\parskip}{0pt} 
\section{\label{sec:level2}THEORETICAL FRAMEWORK FOR $\omega$ MESONS AROUND 2 GeV}

The spectrum of $\omega$ meson states in the 2.0 GeV region presents a complex and poorly understood landscape, as depicted in Fig.~\ref{fig:masses around 2.2}. To elucidate their underlying structures, a detailed theoretical investigation is essential. In this section, we employ the Modified Godfrey-Isgur (MGI) model \cite{Song:2015nia} to calculate the mass spectrum of the light unflavored vector meson family. Beyond mass predictions, the MGI model also provides the spatial wave functions, which serve as critical inputs for subsequent computations of two-body OZI-allowed strong decay widths using the Quark Pair Creation (QPC) model~\cite{Anisovich:2005wf,Roberts:1992js,Blundell:1996as}.

\subsection{The Modified Godfrey–Isgur model}

Potential models have been successfully applied to describe the spectroscopy of heavy quarkonium systems, where the constituent quarks are sufficiently massive to warrant a non-relativistic treatment, as exemplified by the Cornell potential model \cite{Eichten:1974af,Eichten:1978tg,Eichten:1979ms}. {In such systems, the meson mass primarily arises from the interaction between the constituent quarks.} {However, this approach is no longer adequate for light meson systems, since relativistic effects are significant in such cases.} To address this, Godfrey and Isgur developed a relativized quark model in 1985 \cite{Godfrey:1985xj}, which builds upon the Cornell potential while incorporating a comprehensive set of relativistic corrections.

The Hamiltonian describing the interaction between a quark and an antiquark in the MGI model is written as
\begin{equation}
    \begin{aligned}
    {\tilde{H}}={\sqrt{m_{1}^2+p^2}}+{\sqrt{m_{2}^2+p^2}}+{\tilde{V}}_{eff}(p,r),
    \end{aligned}
\end{equation}
where $m_1$ and $m_2$ denote the quark and antiquark masses, respectively. The effective potential takes the form \cite{Song:2015nia}
\begin{equation}
\begin{aligned}
    {\tilde{V}}_{eff}(p,r)= {\tilde{H}}_{OGE}+{\tilde{H}}_{conf} ,
\end{aligned}
\end{equation}
with  ${\tilde{H}}_{OGE}$ representing the one-gluon-exchange potential at short distances, and ${\tilde{H}}_{conf}$ corresponding to the long-range confinement part.

To account for short-distance singularities, a Gaussian smearing transformation is introduced,
\begin{equation}
\begin{aligned}
{\tilde{V}}_{eff}(p,r)= \int d^3r' \rho_{ij}(r-r')\,{V_{eff}}(p,r'),
\end{aligned}
\end{equation}
where the smearing function is defined as
\begin{equation}
\begin{aligned}
{\rho}_{ij}(r-r')=\frac{\sigma_{ij}^3}{\pi^{3/2}}\,e^{-{\sigma_{ij}^2}(r-r')^2}.
\end{aligned}
\end{equation}

For the confinement part, the linear term in the GI model is modified by the screening effect\cite{Song:2015nia} as
\begin{equation}
\begin{aligned}
br \to \frac{b}{\mu}(1-e^{-\mu r})\,,
\end{aligned}
\end{equation}
where $b$ denotes the string tension and $\mu$ is the parameter controlling the screening strength.
The masses of the $\omega$ mesons obtained under this model are listed in Table.~\ref{tab:omega-spectrum}.
\begin{table}[htbp]
\centering
\captionsetup{justification=raggedright, singlelinecheck=false} 
\renewcommand{\arraystretch}{1.2} 
\begin{tabular}{ccc}
\hline\hline
State & $M_{\text{theo}}$ (GeV) & Candidate \\
\hline
$\omega(1S)$ & 0.775 & $\omega(782)$ \\
$\omega(2S)$ & 1.413 & $\omega(1420)$ \\
$\omega(3S)$ & 1.862 & $\omega(1960)$ \\
$\omega(4S)$ & 2.180 & $X(2240)$, $\omega(2290)$, $\omega(2330)$ \\
$\omega(1D)$ & 1.633 & $\omega(1650)$ \\
$\omega(2D)$ & 2.003 &  \\
$\omega(3D)$ & 2.283 & $\omega(2205)$ \\
\hline\hline
\end{tabular}
\caption{Theoretical masses of $\omega$ states and their experimental candidates.}
\label{tab:omega-spectrum}
\end{table}

\subsection{Quark pair creation model}

The QPC model provides a successful description for most OZI-allowed hadronic decays and is thus widely employed in studies of strong decay processes of hadrons~\cite{Yu:2011ta,Wang:2012wa,He:2013ttg,Ye:2012gu,Chen:2015iqa,Pan:2016bac,Liu:2010tr,Li:2008mza,Li:2008et,Li:2008we,Guo:2019wpx,Wang:2024yvo,Wang:2021abg,Wang:2019qyy,Wang:2017iai}. To calculate these widths for the $\omega$ mesons, we employ the QPC model \cite{Micu:1968mk,yaouanc1977}, which provides a framework for evaluating hadronic transition amplitudes.

\begin{figure}[htbp]
  \centering
  \includegraphics[width=\columnwidth]{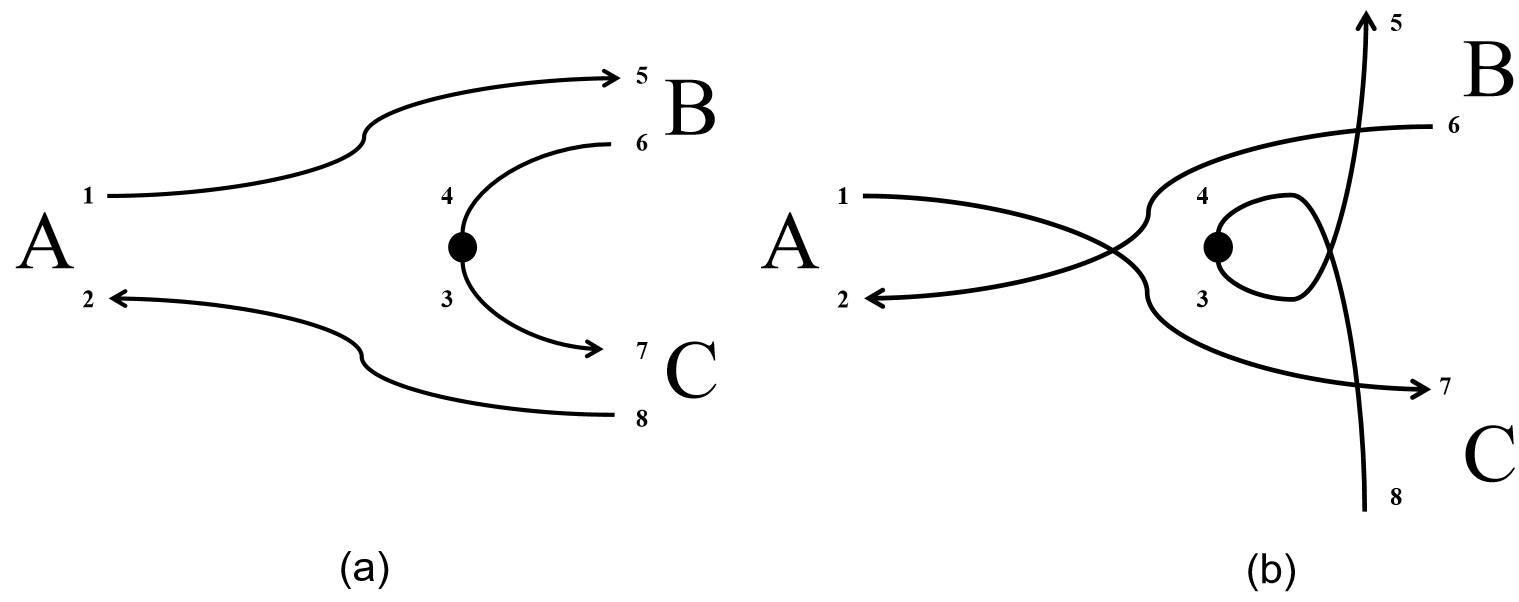} 
  \caption{{The decay diagrams of} $A \to B+C$ mediated by QPC, there are two possible forms. Generally, only one of them contributes to the amplitude.}
  \label{fig:QPC}
\end{figure}

In the QPC model, a decay process $A \rightarrow B+C$ proceeds via the creation of a quark-antiquark pair from the vacuum with the quantum numbers $J^{PC}=0^{++}$. This created pair, in combination with the initial quark and antiquark within the parent meson, subsequently hadronizes into the two outgoing mesons. The decay amplitude is given by
\begin{equation}
    \begin{aligned}
    \langle BC\vert {\cal T}\vert A \rangle = \delta^3(\mathbf{P}_B + \mathbf{P}_C) \times\cal M^{M_{J_A} M_{J_B} M_{J_C}} \mathbf{P}
    \end{aligned}
\end{equation}
where the transition operator $\cal T$ corresponds to the creation of a quark-antiquark pair from the vacuum. The two possible topological diagrams contributing to this amplitude are shown in FIg.~\ref{fig:QPC} \cite{Micu:1968mk,yaouanc1977}.

For a given particle A, the mock state wave function is defined as 
\begingroup
\setlength{\arraycolsep}{2pt}  
\setlength{\jot}{2pt}          
\begin{equation}
\begin{aligned}
&\bigl|A\!\bigl(n_A^{2S_A+1} L_A\, J_A M_{J_A}\bigr)\!(\bm P_A)\bigr\rangle\\
&\equiv \sum_{M_{L_A},\,M_{S_A}}
   \langle L_A M_{L_A}\, S_A M_{S_A}\mid J_A M_{J_A}\rangle \\
&\quad \times \sqrt{2E_A}
   \!\int d^{3}\bm p_A\;
   \psi_{n_A L_A M_{L_A}}(\bm p_A)\,
   \chi_{S_A M_{S_A}}\,
   \phi_A^{12}\,\omega_A^{12} \\
&\quad \times
   \Bigl|\, q_1\!\bigl(\tfrac{m_1}{m_1+m_2}\bm P_A+\bm p_A\bigr)\,
           \bar q_2\!\bigl(\tfrac{m_2}{m_1+m_2}\bm P_A+\bm p_A\bigr)
   \Bigr\rangle .
\end{aligned}
\end{equation}
\endgroup
Here, $m_1$ and $m_2$ are the masses of the quark $q_1$ and antiquark $\bar{q}_2$, respectively, $n_A$  is the radial quantum number, $S_A$ and $L_A$ are the total spin and orbital angular momentum between $q_1$ and $\bar{q}_2$, $J_A=S_A + L_A$ is the total angular momentum, $\mathbf{P_A}=\mathbf{p}_1 + \mathbf{p}_2$ and $E_A$ are the center-of-mass momentum and energy, and $\mathbf{p_A}= \frac{m_1 \mathbf{p_1} - m_2 \mathbf{p_2}}{m_1+m_2}$ is the relative momentum between $q_1$ and $\bar{q}_2$. The terms $\chi , \phi, \omega$, and $\psi$ denote the spin, flavor, color, and spatial wave functions of a meson A , respectively.

The total decay width in the center-of-mass frame is given by
\begin{equation}
    \begin{aligned}
        \Gamma_{A \rightarrow BC } = \frac{\pi}{4} \frac{\vert \mathbf{P} \vert}{m_A^2} \sum_{J,L} \vert \cal M ^{\mathrm{JL} } \mathrm{(} \mathbf{P} \mathrm{)}\vert^{\mathrm{2}}, 
    \end{aligned}
\end{equation}
where $\mathbf{P} = \mathbf{P}_B = -\mathbf{P}_C$, and  are the relative orbital and total angular momenta between the final states. The partial wave amplitude $\cal M ^{\mathrm{JL} } \mathrm{(} \mathbf{P} \mathrm{)}$ is related to the helicity amplitude $\cal M ^{\mathrm{M_{J_A}M_{J_B}M_{J_C}}} \mathrm{(}\mathbf{P}\mathrm{)}$ via the Jacob-Wick formula \cite{Jacob:1959at}.

In the center-of-mass frame, the helicity amplitude $\cal M ^{\mathrm{M_{J_A}M_{J_B}M_{J_C}}} \mathrm{(}\mathbf{P}\mathrm{)}$ is expressed as
\begin{widetext}
\begin{equation}
\begin{aligned}
&\mathcal{M}^{M_{J_A} M_{J_B} M_{J_C}}(\bm P)\\
&= \gamma \sqrt{8E_AE_BE_C}\,
   \sum\limits_{M_{L_A},\,M_{S_A},\,M_{L_B},\,M_{S_B},\,M_{L_C},\,M_{S_C}}
   \langle L_A M_{L_A}\, S_A M_{S_A} \mid J_A M_{J_A} \rangle
   \langle L_B M_{L_B}\, S_B M_{S_B} \mid J_B M_{J_B} \rangle \\
&\quad \times\,
   \langle L_C M_{L_C}\, S_C M_{S_C} \mid J_C M_{J_C} \rangle\,
   \langle 1 m\,\, 1{-}m \mid 00 \rangle\,
   \big\langle \chi^{14}_{S_B M_{S_B}} \chi^{32}_{S_C M_{S_C}}
   \big| \chi^{12}_{S_A M_{S_A}} \chi^{34}_{1-m} \big\rangle \\
&\quad \times\,
   \big[\,\langle \phi_B^{14}\phi_C^{32} \mid \phi_A^{12}\phi_0^{34} \rangle\,
          \mathcal{I}(\bm P,m_2,m_1,m_3)
   + (-1)^{1+S_A+S_B+S_C}\,
     \langle \phi_B^{32}\phi_C^{14} \mid \phi_A^{12}\phi_0^{34} \rangle\,
     \mathcal{I}(-\bm P,m_2,m_1,m_3)\,\big],
\end{aligned}
\end{equation}
\end{widetext}
where $\gamma$ is a universal constant reflecting the creation strength of a quark pair from the vacuum, which {can} be determined from experimental data. Note that the strength for $s \bar{s}$ pair creation differs from that of $u \bar{u} + d \bar{d}$ with the relation $\gamma_s = \gamma_u /\sqrt{3}$ \cite{yaouanc1977}. The flavor wave function of the created pair is denoted by $\phi_0$, and $m_3$ is the mass of the created quark $q_3$.

The momentum-space integral $\cal I$($\mathbf{P} , m_2, m_1, m_3$) is defined as
\begin{equation}
\begin{aligned}
\mathcal{I}(\bm P, m_2, m_1, m_3)
&= \int d^{3}\bm p\;
   \psi^{*}_{n_B L_B M_{L_B}}\!\left(
      \frac{m_3}{m_1+m_3}\bm P + \bm p
   \right) \\
&\quad \times
   \psi^{*}_{n_C L_C M_{L_C}}\!\left(
      \frac{m_3}{m_2+m_3}\bm P + \bm p
   \right) \\
&\quad \times
   \psi_{n_A L_A M_{L_A}}\!\left(\bm P + \bm p\right)\,
   {\cal Y}_{\mathrm{1}}^{\mathrm{m}}(\mathbf{p}),
\end{aligned}
\end{equation}
where ${\cal Y}_{\mathrm{1}}^{\mathrm{m}}(\mathbf{p})$ is a solid harmonic polynomial.

This formulation provides a systematic approach for computing strong decay amplitudes. Utilizing the MGI model for masses and wave functions, and the QPC model for decays, we have calculated the OZI-allowed two-body strong decays for the relevant $\omega$ mesons. The theoretical results for the $\omega(4S)$ and $\omega(3D)$ states, which are crucial for the subsequent analysis of the $e^+ e^- \rightarrow b_1(1235) \pi$ cross section, are summarized in Fig.~\ref{fig:ratio}.
\begin{figure}[htbp]
\centering
\includegraphics[width=\linewidth]{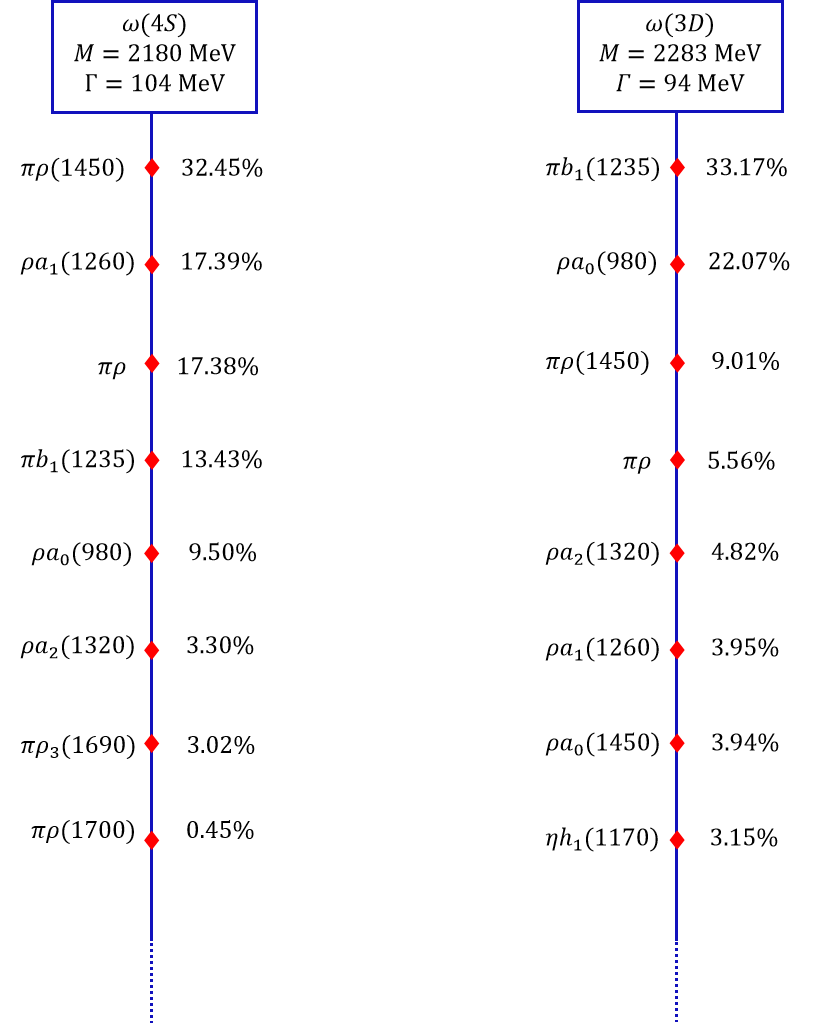}
\caption{\label{fig:ratio} The mass, total width, and the branching ratios of the OZI-allowed strong decays of the $\omega(4S)$ and $\omega(3D)$ states.}
\label{fig:ratio}%
\end{figure}

\vspace{2ex}
\section{\label{sec:level3} DEPICTING THE CROSS SECTION OF $e^+ e^-  \rightarrow b_1(1235) \pi$ AROUND 2 GeV}
\vspace{2ex}

\begin{figure}[htbp]
  \centering
  \includegraphics[width=\columnwidth]{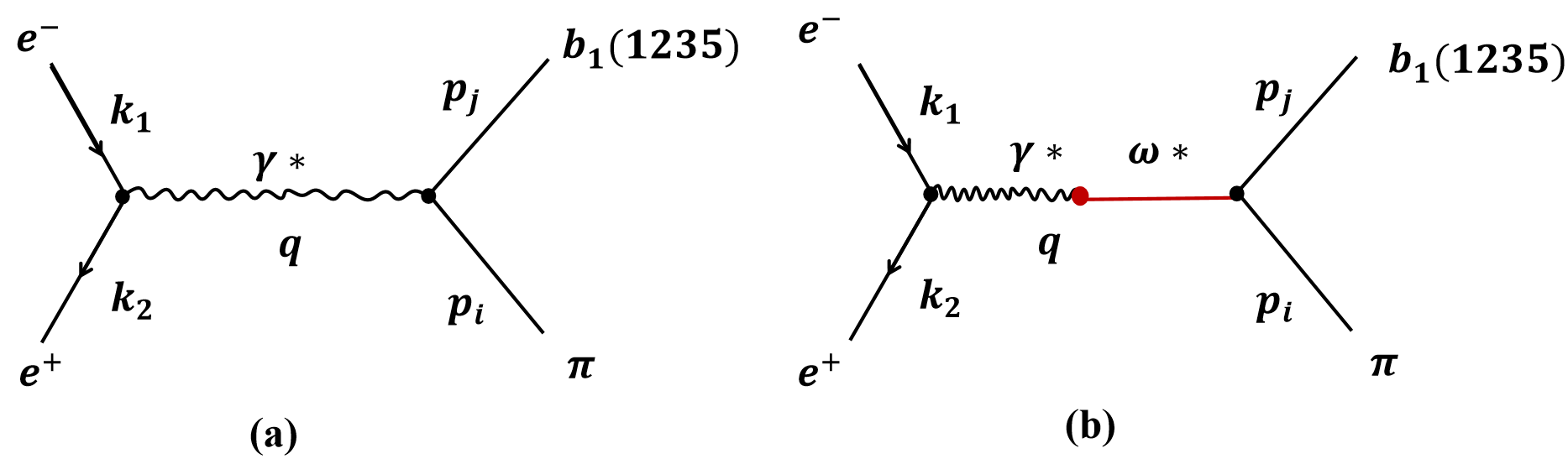} 
  \caption{The schematic diagrams depicting the reaction $e^+ e^-  \rightarrow b_1(1235) \pi$ . Here, diagram (a) is the virtual photon directly coupling with final states, while diagram (b) is due to the intermediate state $\omega^*$ contribution, where $\omega^*$ denotes the $\omega(4S) /\omega(3D)$.}
  \label{fig:feynman}
\end{figure}

This section details the calculation of the cross section for the $e^+ e^-  \rightarrow b_1(1235) \pi$ process around 2 GeV. The conservation of $G$-parity in this reaction suggests that the intermediate resonant state is likely an excited $\omega$ meson. As illustrated in Fig~\ref{fig:feynman}, two primary mechanisms contribute to the cross section: the direct production via a virtual photon (Fig~\ref{fig:feynman}(a)), and the resonant production through intermediate $\omega^*$ states, specifically the $\omega(4S)$ and $\omega(3D)$ (Fig~\ref{fig:feynman}(b)).

To compute these contributions, we employ an effective Lagrangian approach. The relevant interactions are described by the following Lagrangians\cite{Bauer:1975bv,Bauer:1975bw,Kaymakcalan:1983qq,Lin:1999ad,Oh:2000qr,Zhou:2022ark}
\begin{equation}
    {\cal L}_{\gamma \cal V}= \frac{-e m_{\cal V}^2}{f_{\cal V}} {\cal V}_{\mu} {A_\mu},
\end{equation}
\begingroup
\setlength{\abovedisplayskip}{6pt}
\setlength{\belowdisplayskip}{6pt}
\begin{equation}
    {\cal L}_{\gamma {\cal V}{\cal P}}=
    {g}_{\gamma {\cal V}{\cal P}}
    {\varepsilon}_{\mu \nu\alpha\beta}
    {\partial}^{\mu} A^\nu
    {\partial}^{\alpha} {\cal V}^\beta {\cal P},
\end{equation}
\vspace{-0.35\baselineskip}
\begin{equation}
    {\cal L}_{{\cal V} {\cal V}{\cal P}}=
    {g}_{{\cal V} {\cal V}{\cal P}}
    {\varepsilon}_{\mu \nu\alpha\beta}
    {\partial}^{\mu} {\cal V}^\nu
    {\partial}^{\alpha} {\cal V}^\beta {\cal P},
\end{equation}
where $\cal V$ and $\cal P$ stand for the vector and pseudoscalar meson fields, respectively.

The amplitudes corresponding to the diagrams in Fig.~\ref{fig:feynman}(a) and \st{in} Fig.~\ref{fig:feynman}(b) can be expressed as
\vspace{0.5ex}
\begin{equation}
\begin{aligned}
{\cal{M}_{\mathrm{Dir}}} = {\bar v}\mathrm{(}k_{1}\mathrm{)}\,\mathrm{(}ie\gamma_{\mu}\mathrm{)}\,{u}\mathrm{(}k_{2}\mathrm{)}
\mathrm{(}-\frac{g^{\mu\nu}}{s}\mathrm{)}\\
{\times}
\epsilon_{\rho\nu\alpha\beta}{q}^{\rho}\,{p}_{j}^{\alpha}\varepsilon^{*\beta}(p_{j}) {\cal F}(s),
\end{aligned}
\end{equation}
\begin{equation}
\begin{aligned}
\mathcal{M}_{\omega^*}
&= g_{\omega^{*}( b_{1}(1235)\pi)}\;
\bar v(k_{1})\,(ie\gamma_{\mu})\,u(k_{2})
\\
&\times\left(-\frac{g^{\mu\nu}}{s}\right)\left(-\frac{e\,m_{\omega^{*}}^{2}}{f_{\omega^{*}}}\right)
\frac{\tilde g_{\mu\nu}}{\,s-m_{\omega^{*}}^{2}+ i m_{\omega^{*}}\Gamma_{\omega^{*}}\,}
\\
&\times\;
\epsilon_{\delta\rho\alpha\beta}\,
q^{\delta}\,p_{j}^{\alpha}\,\varepsilon^{*\beta}(p_{j}),
\end{aligned}
\end{equation}
respectively, where ${k_1}\, ,{k_2}\,,{p_i}$ and ${p_j}$ are four-momentum of ${e^+},{e^-},\pi$, and ${b_1{(1235)}}$, respectively, and ${q=k_1+k_2}={p_i+p_j}$. ${\cal F}{(s)}=a \exp{(-b\sqrt{s}-\sum_{i} m_i)}$ denotes the form factor, where $a$ and $b$ are free parameters, which can be determined by fitting experimental data, and $\sum_{i} m_i$ is the sum for the masses of the final particles. And $\tilde{g}_{\mu\nu} = -g_{\mu\nu}+q^{\mu}q^{\nu}/q^2$. 
The parameter $f_{\omega^*} $ can be related to the dilepton decay  width of $\omega^*$ by
\begin{equation}
   \Gamma_{\omega^* \rightarrow e^+e^-}=\frac{e^4\,m_{\omega^*}}{12\pi f_{\omega^*}^2}. 
   \label{eq:ratio}
\end{equation}

With the above amplitudes, the differential cross sections of these processes can be calculated by
\begin{align}
{d\sigma}
&= \frac{1}{32\pi s}\,
   \frac{\lvert \vec{p}_{j,\mathrm{cm}}\rvert}{\lvert \vec{k}_{1,\mathrm{cm}}\rvert}\,
   \overline{\lvert \mathcal{M}_{\mathrm{Total}}\rvert^{2}}{d\cos\theta},
\end{align}
where $\theta$ is the scattering angle of an outgoing $\omega$, relative to the direction of the electron beam in the center-of-mass frame, while ${\vec{k}_{1,\mathrm{cm}}}$ and $\vec{p}_{j,\mathrm{cm}}$ are the three-momentum of the electron and $b_1(1235)$ in the final states. The overline in $\overline{\vert{\cal M}_{Total}\vert ^{2}}$ indicates the average over the polarization of $e^{+}e^{-}$ in the initial states and the sum over the polarization of $b_1(1235)\pi$ in the final states. In addition, ${\cal M}_{Total}$ is the total amplitude of the $e^+ e^-  \rightarrow b_1(1235) \pi$ process, which is expressed as follows
\begin{equation}
{\cal M}_{Total}={\cal M}_{Dir}+{\sum_{R}{\cal M}_{R}}e^{i \phi_{R}},
\end{equation}
where $R$ stands for different intermediate $\omega^*$ meson states, and $\phi_{R}$ denotes the phase angle of amplitudes of the direct annihilation and the intermediate excited $\omega$ meson contribution.

It should be noted that the dilepton width $\Gamma_{e^+ e^-}$ presented in Eq.~\ref{eq:ratio}  is evaluated using the relation:
\begin{equation}
\Gamma_{e^+ e^-}=\frac{4\pi}{3} {\alpha}^2 m_{\omega^*} {\cal M}_{\omega^*}^2,
\label{eq:ee}
\end{equation}
where ${\cal M}_{\omega^*}$ denotes decay amplitude, which is defined as ${\cal M}_{\omega^*}=\sqrt{2/3}V_{\omega^*}$ and ${\cal M}_{\omega^*}=\sqrt{4/27}V_{\omega^*}^{'}$ for S–wave and  D–wave $\omega$ mesons, respectively. The factors $V_{\omega^*}$ and $V_{\omega^*}^{'}$ are given by:
\begin{equation}
\begin{aligned}
V_{\omega^*} &= m_{\omega^*}^{-2}\,\tilde{m}_{\omega^*}^{1/2}\,(2\pi)^{-3/2}
\int d^3p \, (4\pi)^{-1/2}\,\phi_{\omega^*}(p)\\
&\times p^2
\left(\frac{m_1 m_2}{E_1 E_2}\right)^{1/2} \,,
\end{aligned}
\end{equation}

\begin{equation}
\begin{aligned}
V'_{\omega^*} &= m_{\omega^*}^{-2}\,\tilde{m}_{\omega^*}^{1/2}\,(2\pi)^{-3/2}
\int d^3p (4\pi)^{-1/2}\,\phi_{\omega^*}(p)\\
&\times p^2
\left(\frac{m_1 m_2}{E_1 E_2}\right)^{1/2}
\left(\frac{p}{E_1}\right)^2 \, .
\end{aligned}
\end{equation}
with constituent quark mass $m_1=m_2=0.22$ GeV \cite{Godfrey:1985xj}, and $\tilde{m}_{\omega^*}=2{\int d^3p\,E\vert \phi_{\omega^*}(p)\vert^2 p^2}$. The spatial wave fanction $\phi_{\omega^*}(p)$ is described by the simple harmonic oscillator wave function with size parameter $\beta=0.41$ GeV. Using this formalism, we obtain $\Gamma_{\omega(4S)\to e^+e^-} = 7\,\mathrm{eV}$ and $\Gamma_{\omega(3D)\to e^+e^-} = 1.8\,\mathrm{eV}$, which through Eq.~(\ref{eq:ratio}) yield the decay canstants $f_{\omega(4S)}=264$ and $f_{\omega(3D)}=532$. These calculated values are consistent with those extracted from the cross section fit, thereby reinforcing the consistency of our resonance interpretation.

\vspace{2ex}
\section{\label{sec:level4} NUMERICAL ANALYSIS AND RESULTS}
\vspace{2ex}

This section presents the numerical analysis of the Born cross section for $e^+ e^-  \rightarrow b_1(1235) \pi$ measured by the BESIII Collaboration \cite{BESIII:2022yzp}, where an event accumulation near 2.2 GeV exists in the $ b_1(1235) \pi$ \cite{BESIII:2022yzp} invariant mass spectrum. Our objective is to quantitatively interpret the enhancement structures near 2.2 GeV by extracting the contributions of the $\omega(4S)$ and $\omega(3D)$ resonances. Guided by the theoretical predictions on the mass spectrum and decay properties from Sec.~\ref{sec:level2}, we perform a fit to the experimental data within the framework developed in Sec.~\ref{sec:level3}.

\renewcommand{\arraystretch}{1.4} 
\begin{table}[htbp]
\centering
\captionsetup{justification=raggedright, singlelinecheck=false} 
\begin{tabular*}{\columnwidth}{@{\extracolsep{\fill}}lc}
\hline\hline
Parameters & Values \\
\hline
$a(\mathrm{GeV}^{-1})$ & $0.85 \pm 0.03$ \\
$b(\mathrm{GeV}^{-1})$ & $1.12 \pm 0.02$ \\
$\phi_{\omega(3D)}\,(\mathrm{rad})$ & $1.62 \pm 0.13$ \\
$\phi_{\omega(4S)}\,(\mathrm{rad})$ & $1.48 \pm 0.07$ \\
$g_{\omega(4S)b_1(1235)\pi}\,(\mathrm{GeV})$ & $0.0354 \pm 0.0012$ \\
$g_{\omega(3D)b_1(1235)\pi}\,(\mathrm{GeV})$ & $0.0634 \pm 0.0053$ \\
\hline\hline
\end{tabular*}
\caption{The parameters obtained by fitting the cross section of $e^+ e^-  \rightarrow b_1(1235) \pi$ measured by BESIII \cite{BESIII:2022yzp}, and the $\chi^2$/n.d.f value is 1.28 for this fitting.}
\label{tab:fitparams}
\end{table}

\begin{figure}[htbp]    
\centering    
\includegraphics[width=\columnwidth]{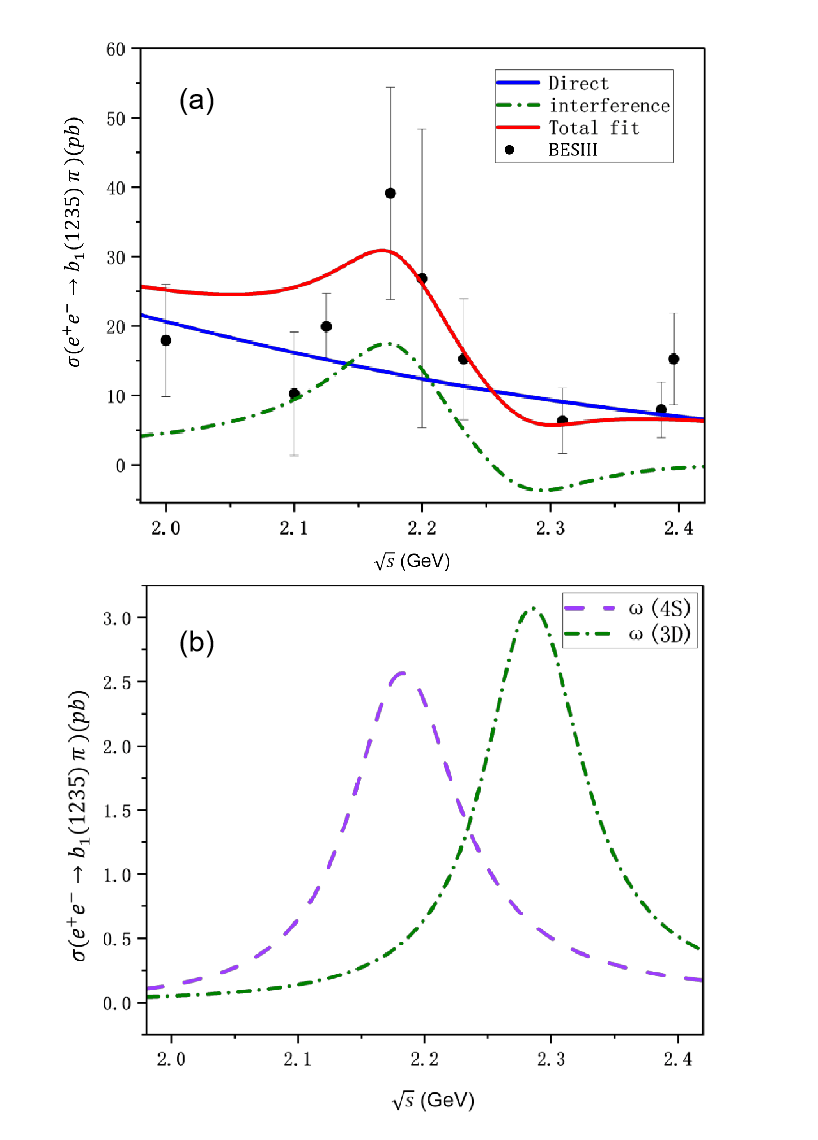} 
    \caption{The fitted result of the experimental data of the Born cross sections of $e^+ e^-  \rightarrow b_1(1235) \pi$ (a), and the calculated cross sections of $e^+ e^-  \rightarrow  \omega(4S)/\omega(3D) \rightarrow b_1(1235) \pi$ \,(b).}    
\label{fig:b1}
\end{figure}

The fitting procedure yields the parameters summarized in Table.~\ref{tab:fitparams}. The satisfactory fit quality, with $\chi^2/\text{n.d.f})=1.28$ across the entire energy range, indicates that the two-resonance interference scenario provides an adequate description of the data. This value, falling within the ideal range of 1 $\sim$ 1.5, confirms that our model captures the core features of the data without overfitting or underfitting.

The resulting fit is presented in  Fig.~\ref{fig:b1}(a), which shows good agreement with the experimental cross section. Our analysis reveals that both the $\omega(4S)$ and $\omega(3D)$ states contribute significantly and with comparable magnitude to the observed cross section. This finding resolves the previously noted discrepancy in resonance parameters. Specifically, the $\omega(4S)$ resonance produces a prominent peak around 2.2 GeV, while the $\omega(3D)$ generates a broader enhancement centered near 2.3 GeV. The individual resonant cross sections are shown in Fig.~\ref{fig:b1}(b). The interference between these states, characterized by the fitted phase angles $\phi_R$, is essential for reproducing the detailed lineshape of the enhancement.

In conclusion, the enhancement structure observed in $e^+ e^-  \rightarrow b_1(1235) \pi$ near 2.2 GeV can be attributed to the combined contributions of the $\omega(4S)$ and $\omega(3D)$ resonances. This study provides valuable information for establishing the $\omega$ meson family in this mess region.

\section{\label{sec:level5} SUMMARY}
\vspace{2ex}

The process $e^+ e^- \rightarrow b_1(1235) \pi$ provides a unique laboratory for probing the higher-mass region of the light vector meson spectrum, particularly due to the constraint of $G$-parity conservation, which favors the production of $\omega$-like states. This potential is now realizable with the high-precision measurement of the corresponding cross sections around 2.2 GeV by the BESIII Collaboration, which motivates our present work. To decipher the underlying structure of the observed enhancement, we have conducted a combined analysis that integrates theoretical spectroscopy with these experimental data.

Our theoretical framework is anchored in the mass spectrum and decay properties predicted by the MGI model, while the decay widths are computed using the QPC model. The application of this framework to the experimental cross sections leads to a pivotal finding: the enhancement near 2.2 GeV does not originate from a single resonance. Instead, it is predominantly generated by the significant constructive interference between the $\omega(4S)$ and $\omega(3D)$ states. Crucially, our analysis indicates that both states contribute to the 
$b_1(1235)\pi$ channel with comparable strength. This interpretation, supported by the fitted resonance parameters and phase angles, successfully resolves previous experimental discrepancies and aligns with theoretical expectations.

Consequently, this study establishes a coherent description of the vector enhancement near 2.2 GeV, providing a robust framework for identifying higher radial and orbital excitations in the $\omega$ meson family. It marks a significant advancement in mapping the light-flavor meson spectrum and highlights the essential synergy between precision experiments and theoretical models in driving progress in hadronic physics.

\vspace{2ex}
\begin{acknowledgments}
This work is supported by  the National Natural Science Foundation of China under Grant No.~12405104, the Natural Science Foundation of Hebei Province under Grant No.~A2022203026, the Higher Education Science and Technology Program of Hebei Province under Contract No.~BJK2024176, and the Research and Cultivation Project of Yanshan University under Contract No.~2023LGQN010.
\end{acknowledgments}

\nocite{*}

\bibliography{apssamp}

\end{document}